# The impact of morphological structure and flexo-chemical strains on the electric transport mechanisms in the molybdenum-disulfide-oxide nanoflakes

O. S. Pylypchuk[1], V. V. Vainberg[1*], V. N. Poroshin[1], A. V. Terebilenko[2], A. S. Nikolenko[3], V. I. Popenko[3], A. S. Tolochko[1], M. V. Olenchuk[1], O. Bezkrovnyi[4], G. I. Dovbeshko[1], T. Sabov[3], B. M. Romanyuk[3], S. V. Kolotilov[2†] and A. N. Morozovska[1‡]

[1] Institute of Physics, National Academy of Sciences of Ukraine, 46, pr. Nauky, 03028 Kyiv, Ukraine

[2] L.V. Pisarzhevskii Institute of Physical Chemistry, National Academy of Sciences of Ukraine, 31, pr. Nauky, 03028 Kyiv, Ukraine

[3] Lashkarev Institute of Semiconductor Physics, National Academy of Sciences of Ukraine, 41, pr. Nauky, 03028 Kyiv, Ukraine

[4] W. Trzebiatowski Institute of Low Temperature and Structure Research, Polish Academy of Sciences, 50-422 Wroclaw, Poland

## ABSTRACT

Electric conduction mechanisms are studied in the pressed nanoflake powder of the molybdenum-disulfide-oxide ($MoS_xO_y$) depending on their content and structure. The $MoS_xO_y$ nanoflakes were prepared by reaction of $(NH_4)_6Mo_7O_{24}$ with thiourea in aqueous solution followed by aerial oxidation. The sintered nanoflakes are 10-20 nm thick and self-assembled in the "nanoflower"-shape aggregates forming powder particles. The chemical composition and structure of the powders were studied by XPS, EDS and Raman spectroscopy, which show that the powders have different chemical composition and structure depending on the preparation conditions. These studies revealed the existence of different forms of $MoS_2$ and its oxides in the powders. These features are impactful on electric transport properties. The current vs voltage (I-V) dependences of the pressed $MoS_xO_y$ nanoflakes reveal hysteresis-like behavior; and their loop width depends on the chemical composition and structure. In the samples with the highest content of Mo in the oxide/sulfoxide form the negative differential conductivity has been observed. The I-V curves of all $MoS_xO_y$ nanoflake samples manifest the three-state resistive switching and the long-lasting transient charge/discharge on switching "on/off" the voltage across the sample, which evidences the role of interface charges in their conductivity. To describe theoretically the observed I-V curves, polar and electric-transport properties of the pressed $MoS_xO_y$ nanoflakes, the Landau-Cahn-Hilliard approach considering flexo-chemical field has been used. The revealed in experiment and explained theoretically features of resistive switching and charge accumulation look promising for applications in memristors and high-performance supercapacitors.

---

[*] corresponding author, e-mail: viktor.vainberg@gmail.com
[†] corresponding author, e-mail: s.v.kolotilov@gmail.com
[‡] corresponding author, e-mail: anna.n.morozovska@gmail.com (submitting author)

## 1. INTRODUCTION

Semiconducting low-dimensional (LD) transition metal dichalcogenides (TMD), in the form of flexible ultrathin films [1], nanoflakes and their flower-like arrays [2, 3, 4], are promising candidates for nanoelectronics [5, 6, 7], supercapacitors [8], strain- and isotope-engineering [9,10], and flexo-tronics [11]. The possibilities of controlling the structural, polar, and electronic properties of the LD TMD by applying homogeneous compressive or tensile elastic strains [12, 13, 14], and bending deformations [15, 16] have been revealed.

Even in the case of the chemically pure $MoS_2$ nanoflakes ensembles (e.g., in the form of powders) there may emerge that the part of flakes undergoes the phase transition from the semiconducting 2H phase to the metallic 1T' phase. The transition may be caused by intercalative doping and mechanical compression and filament formation in the strong electric field. For example, such switching was observed in the I-V curves of the $MoS_2$ powder starting from relatively small compressions (above 0.2%) and electric fields more than (80 – 100) V/cm [17].

The thermal annealing can significantly influence structural, optic and electronic properties of $MoS_2$ nanosheets [18] and thin films [19]. In particular, the metastable metallic 1T phase emerges with Li intercalation into a few-layer thick $MoS_2$, and the mild annealing at temperatures from 50ºC to 300ºC leads to gradual restoration of the 2H semiconducting phase [20]. The annealing impacts sulfur vacancies and, respectively, electronic transport in $MoS_2$ film [21] and allows defect engineering on the $MoS_2$ surface [22].

Besides the strain- and isotope-engineering of LD TMDs, great attention is paid to studies of complex materials consisting of $MoS_2$ nanopowders mixed with Mo-containing compounds like $MoO_3$ [23, 24] or $Mo_2S_3$ [25, 26]. These efforts are related to the development of clean energy conversion and storage systems, hydrogen production devices, supercapacitors, including fabrication technology, structure features and electric properties of resistive switching devices [27, 28].

Similar to epitaxial elastic strains, the chemical strains [29, 30], induced by impurities, such as elastic defects (elastic dipoles or dilation centers), can also control the electronic properties [31] and local symmetry [32, 33] of LD TMDs. Above a certain deformation threshold, chemical strains can induce ferroelectric-like domain structures and conductive domain walls in the LD-TMD [31-33].

The possibility of controlling the electronic and polar properties of LD TMDs by bending uses the fact that the flexoelectric effect can play a very important role [34]. Flexoelectricity, defined as coupling between the strain gradient (e.g., bending or rippling) and electric polarization [35], can be responsible for many important features in the LD TMD [11, 36]. The flexoelectricity is expected to be extremely strong in sliding van der Waals (vdW) ferroelectrics [37], such as TMD nanoflakes. Flexoelectricity can induce, enhance or reverse the polarization, change domain chirality and influence strongly the domain morphology in versatile ferroic films [38] and nanoparticles [39], including nanoflakes with vdW layered structure [40].

The sliding flexo-ferroelectricity can induce the interfacial charge accumulation in the curved LD layered vdW TMDs [41, 42]. The sliding ferroelectricity has been revealed in artificially stacked 2D TMD [43, 44] with out-of-plane electric polarization originating from the interlayer charge transfer in their stacking. The stack polarization can be reversed across the interlayer sliding with an ultralow barrier [45]. The LD vdW materials, which are free-standing sliding ferroelectrics, typically have high fracture stress and low bending stiffness, enabling their significant out-of-plane curvature, corresponding to the strain gradients of several orders of magnitude higher than those achieved in epitaxial ultra-thin oxide films clamped to rigid substrates [46].

Despite a lot of performed studies on structural and electrophysical properties of the LD TMD, including MoS$_2$, insufficient attention was paid to changing their properties by modifying composition and properties due to oxidation or coupling with oxides in a single structure. Partial oxidation of MoS$_2$ and formation of molybdenum oxides may be an efficient tool for engineering of required properties. Such studies should be useful in development of devices for the resistive switching and energy storage.

In this work we explore the electric transport properties of powders consisting of nanoflakes of the molybdenum sulfide-oxide, which chemical composition can be described by a general formula MoS$_x$O$_y$, with x and y changing in a wide range. The nanoflakes were prepared by the hydrothermal reaction of (NH$_4$)$_6$Mo$_7$O$_{24}$ with thiourea in aqueous solution and synthesized at different temperatures $T_s$ from 130°C to 180°C. The flakes are 10-20 nm thick and self-assembled in nanoflower-like agglomerates forming the powder particles. The preparation method and description of the structural morphology of the nanoflakes studied in this work are given in detail in Ref. [47]. Below we give a brief description for the sake of completeness.

## 2. MORPHOLOGY OF NANOFLAKES

The morphology of the samples studied is classified depending on the synthesis temperature $T_s$. For the sake of convenience, the samples are marked using the abbreviation "MoSOT$_s$", e.g., MoSO160 means MoS$_x$O$_y$ sintered at 160°C. The powders consist of the flower-like conglomerates; the average size of conglomerates gradually increasing from 0.2 to 1 μm with an increase in T$_s$ from 130°C to 180°C [47]. The conglomerates consist of the self-assembled 10-20 nm thick nanoflakes.

The chemical composition of the powders was obtained by EDS. All samples are characterized by a large content of oxygen, which is comparable with that of sulfur. The total amount of the O and S atoms is more than twice as large as compared with that in a pure MoS$_2$. The contents of O, S and Mo in three typical samples, MoSO140, MoSO150 and MoSO160, are listed in **Table 1**. The distribution of chemical elements in the form of certain crystalline phases in the MoS$_x$O$_y$ nanoflakes, was determined from the XPS and Raman spectra. The distribution of chemical elements, typical for the samples MoSO140, MoSO150 and MoSO160, is listed in **Table 2**.

**Table 1.** Elemental content of the MoS$_x$O$_y$ nanoflake powders synthesized at different T$_s$, determined from EDS studies

| Sample | Oxygen, at. % | Sulphur, at.% | Molybdenum, at.% |
|---|---|---|---|
| MoSO140 | ≈42±2.1 | ≈41.2±1 | ≈16.3±1.5 |
| MoSO150 | ≈35.8±2.5 | ≈45.8±2.6 | ≈18.2±0.3 |
| MoSO160 | ≈46.12±0.6 | ≈36.3±0.4 | ≈17.6±1 |

**Table 2.** Distribution of chemical elements in possible compounds, determined from XPS studies

| Sample | Mo in sulfide, % | Mo in Oxide/sulfoxide, % | S(VI) (Sulfate), % |
|---|---|---|---|
| MoSO130 | 74 | 26 | 7.8 |
| MoSO140 | 76, 87 | 24, 13 | 3.0, 2.6 |
| MoSO150 | 48, 34 | 52, 66 | 66.4, 61.2 |
| MoSO160 | 75, 59 | 25, 41 | 2.6, 9 |

The analysis of the Raman spectra, averaged over several points in each sample, enables us to distinguish the following inherent features:

I. The presence of the amorphous phase, which corresponds to the amorphous regions with broad bands.

II. The presence of the MoS$_2$ phase, which spectral bands are characteristic for MoS$_2$ nanostructures.

III. The presence of the crystalline oxide phase, which spectra has narrow bands of crystalline molybdenum oxide (registered for areas in the form of crystallites of a regular shape).

Notably one should distinguish the sample MoSO150 for which the MoS$_2$ phase is not observed. It is important, that a certain oxidation or crystallization to the α-MoO$_3$ phase occurs in the Raman spectra measured at increased intensity of excitation beam for the most samples in amorphous areas. One observes a slight restructuring of the spectrum for the sample MoSO150 in its amorphous areas, probably due to the oxidation. The bands become more pronounced but maintain features typical for the amorphous phase.

To summarize the above characterization results, the sample MoSO150 differs significantly from other samples, which is important for their electric transport features considered below.

## 3. EXPERIMENTAL

Three different kinds of samples were prepared to study electric transport features. In the first kind the powder was tableted as a disk with a diameter of 4 mm and thickness of about 150 μm. These sizes were chosen to provide acceptable limits for measurements of the sample resistance and to provide the sample thickness to be much larger than nanoflakes sizes to avoid the short circuit conditions. The sample was placed between compressing brass plungers in the cylindrical Teflon tube (**Fig. 1(a)**). The metallic plungers serve both as electric contacts and to perform uniaxial compression of the sample in the pressure range of 0.5–2.5 MPa. These samples were used to investigate the transient processes during charge accumulation and discharge after switching on/off the voltage across the sample and to check the impact of applied uniaxial pressure. The second kind is shown in **Fig.1(b)**. The required amount of powder film is applied to a textolite plate with strip metallic contacts. The film is flattened and formed by applying the uniaxial pressure of 2.5 MPa for 10 seconds. The sample resistivity remained after that unchanged during long time within the third digit. This kind of samples was used to measure current-voltage characteristics. The third kind of samples was made similar to the second one, and it had an additional electric contact in the middle between two end contacts (**Fig.1(c)**). This kind was used to study the characteristics of the current contacts and their impact on the measured current-voltage characteristics (I-V curves). Shown in **Fig.1(d)** is the principal electric circuit of electric measurements with the abovementioned three kinds of samples.

The electric voltage was applied to the sample, connected in series with the load resistor, from the software-controlled dc power supply GW Instek PSP-603. The voltage drops across the sample and the load resistor were measured by the digital multimeters Keithley 2000 and recorded by the computer. To record quick transient processes, we used the digital oscilloscope Tektronix TDS1002B and the pulsed voltage supply G5-63. The dielectric characteristics were measured by the LCR-meter LCX200 ROHDE & SCHWARZ.

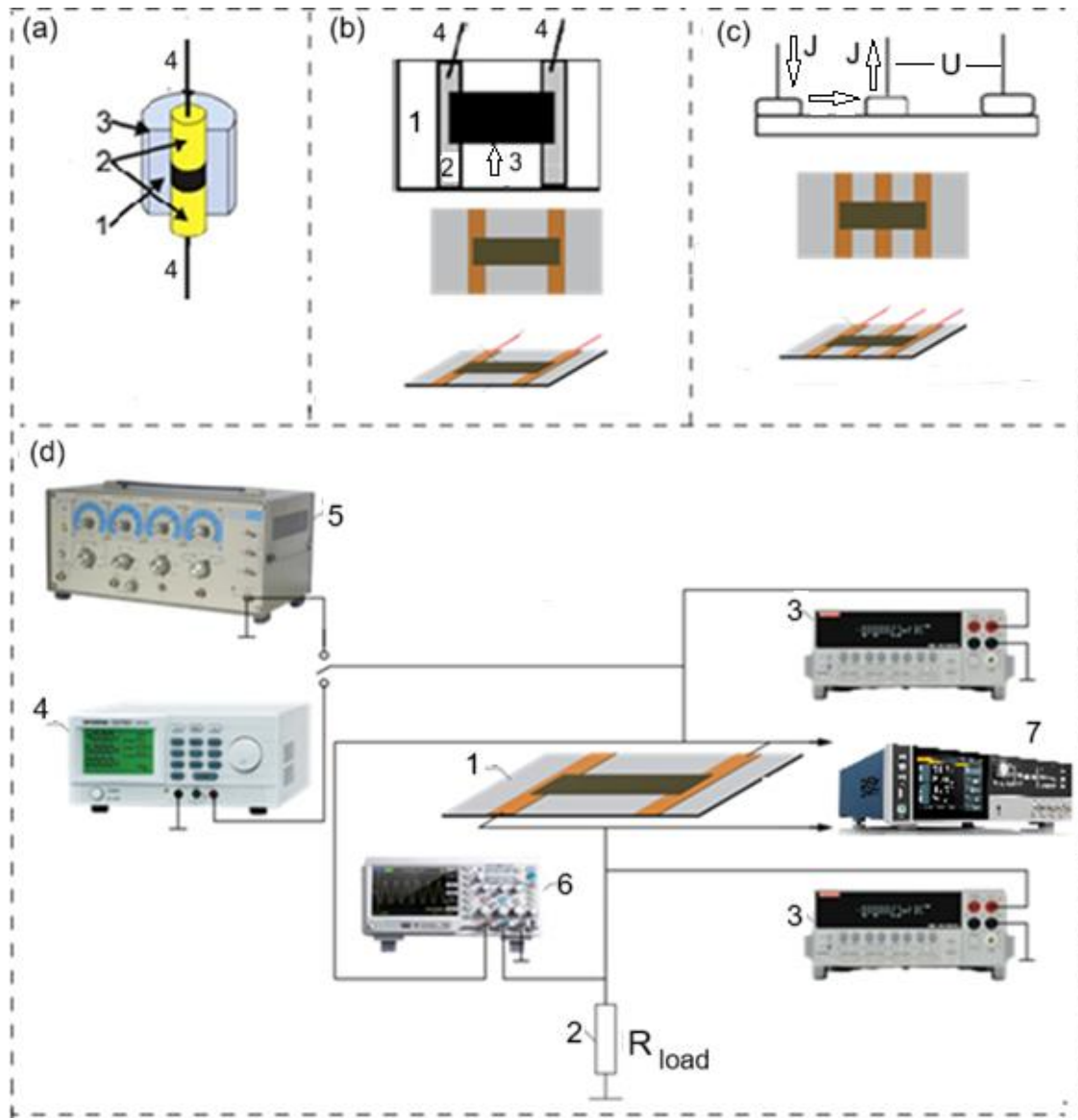

**Figure 1.** Schematic images of the samples. **(a)** The cell with a sample to measure charge/discharge transient characteristics under controlled uniaxial pressure. "1" is the powder disk sample, "2" are the brass plungers, "3" is the Teflon tube, "4" are the electric wires. **(b)** The flat preliminarily compressed powder sample to measure current-voltage characteristics. "1" is the textolite substrate, "2" are the metallized strips of electrical contacts, "3" is a powder sample compressed as a film on the substrate, 4 are the electric wires. **(c)** The schematic view and real image of the sample for investigation of electric contact characteristics. **(d)** The principal electric circuit of electric measurements. 1 is the sample, 2 is the load resistor, 3 are the digital multimeters, 4 is the dc power supply, 5 is the pulsed power supply, 6 is the oscilloscope, 7 is the LCR meter

# 4. RESULTS AND DISCUSSION

## 4.1. Current-voltage characteristics

For electrophysical measurements the $MoS_xO_y$ powder samples were formed under uniaxial pressure of 2.5 MPa. The I-V curves were measured in the direct current (dc) regime with the voltage step of 20 mV and a delay after each step sufficient to stabilize the current through the sample at temperature 295 K. The delay after each step is caused by the long-lasting transient processes after increasing voltage. The I-V curves for the $MoS_xO_y$ powder samples are shown in **Fig. 2**.

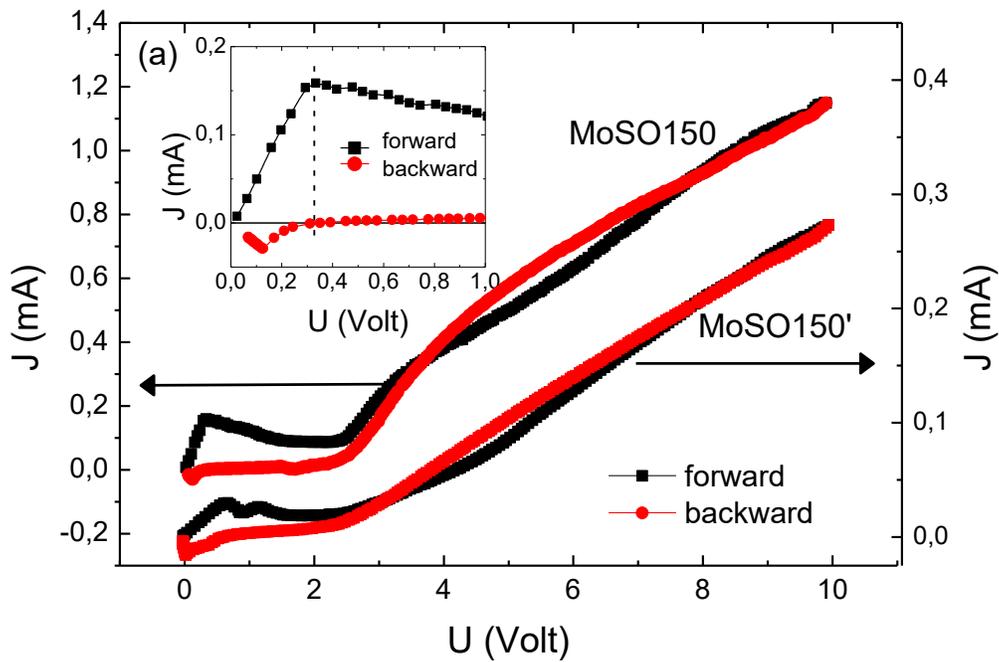

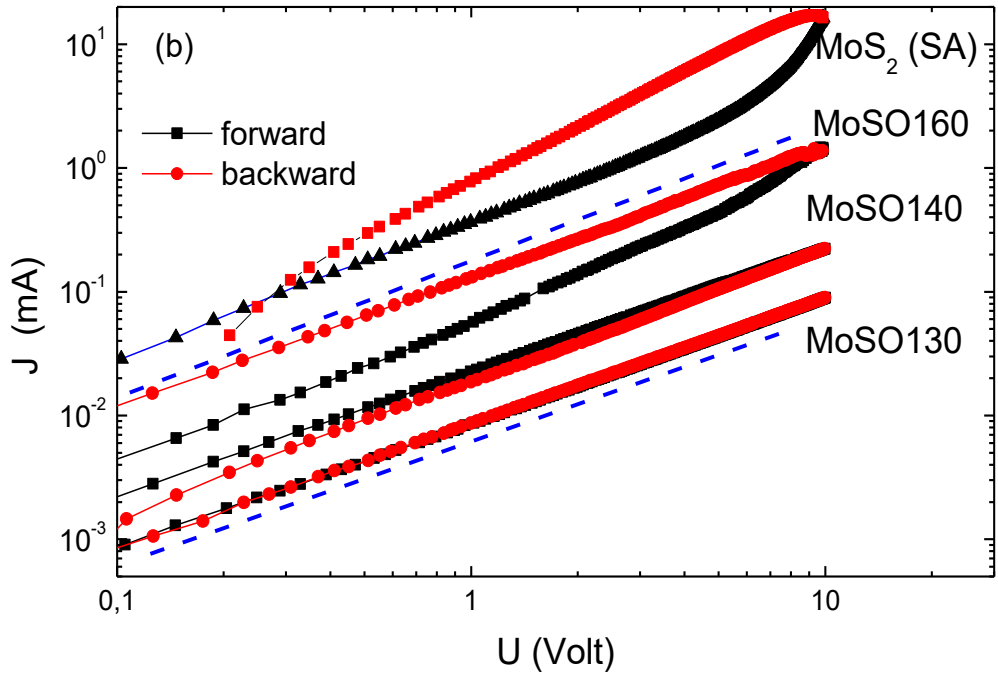

**Figure 2. (a)** Typical I-V curves of 2 MoSO150 samples, which reveal the pronounced NDC effect. Inset: zoomed I-V curves with the NDC part in the forward and backward branches. **(b)** The I-V curves of the MoSO130, MoS140 and MoS160 samples. The blue dashed lines correspond to the ohmic behavior of the I-V curves. The I-V curves of the $MoS_2$ sample pressed from the powder produced by Sigma Aldrich [TM], denoted as $MoS_2$ (SA) are added for comparison.

The I-V curves were measured by the two-contact method. To estimate the contribution of the contact resistance, we performed the measurements of the comtact I-V curves by the method illustrated in **Fig.1(c)**. The current passed through the left and central contacts, and the voltage drop was measured between the central and right contacts. The most crucial impact of the contact resistance on the total I-V curves was observed for the sample MoSO150. The dependence of the contact resistance, $R_c$, and sample resistance, $R_s$, on the voltage across the sample is shown in **Fig. 3** for the sample MoSO150. It is seen that the contact resistance is less by an order of magnitude than the whole sample resistance. At voltage across the sample more than 0.2 V, both $R_c$ and $R_s$ increase synchronously by more than an order of magnitude, which evidences the physical relation between $R_c$ and bulk resistivity and reflects changes in the bulk conduction mechanism. For other samples the ratio of $R_c$ and $R_s$ is less than that for the MoSO150 sample. This means that contact properties do not impact noticeably on the measured I-V curves.

As follows from **Fig. 2**, the samples under study can be divided conditionally by two kinds. This division may be related with the sharp difference in the content of "Mo in the oxide/sulfoxide" phase (see **Table 2**). The sample MoSO150 contains 55-66 % of this phase, that is roughly twice as large as compared to other samples and correspondingly it has twice smaller amount of the "Mo in sulfide" phase, 34-48 % against 74-76 % in other samples. Even more drastic difference between the samples is in the estimate of the "S(VI) (Sulfate)" phase content.

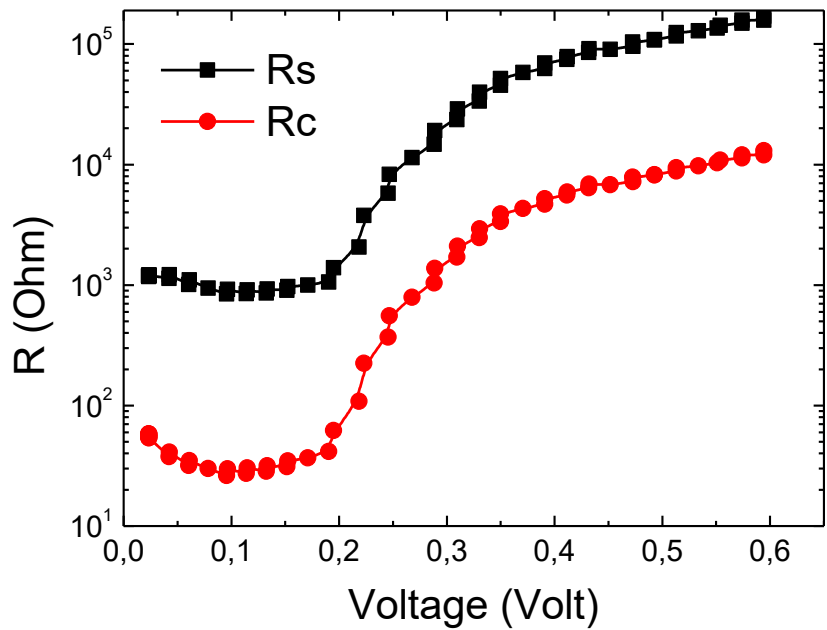

**Figure 3.** Dependence of the contact resistance, $R_c$, and whole sample resistance, $R_s$, on the voltage drop across the sample, measured at room temperature T=293 K.

The consequence of the clearly correlated difference in the chemical content is that the I-V curves of the MoSO150 sample differ from other samples by a pronounced negative differential conductivity (NDC) part, which is absent in other samples. For all samples, as well as for the MoSO150 sample, one can see a pronounced hysteresis loop between the forward and backward (increasing and decreasing applied voltage) branches. Both kinds of I-V curves are well-reproducible. For example, the curves for the MoSO150 sample presented in **Fig. 2(a)** correspond to two different powder samples synthesized in different times. Concerning the samples MoSO130, 140 and 160, their I-V curves have a slightly non-linear (but close to ohmic) behavior with a hysteresis between forward and backward branches, and the hysteresis loop widening with an increase in $T_s$. The I-V curve of the pure $MoS_2$ sample made of the powder produced by Sigma Aldrich [TM] is shown in **Fig. 2(b)** for comparison. All I-V curves in **Fig. 2(b)** look similar, which agrees with the shift of the phase ratio in **Table 2** to the larger content of the "Mo in sulfide" phase. It is worth noting that the MoSO samples in **Fig. 2(b)** have different mutual layouts of the forward and backward branches. At the lower $T_s$ the current in the forward branch is higher than that in the backward branch at the same applied voltage, while they change their layout to the opposite at the higher $T_s$. The similarity of the I-V curves for the MoSO130, MoSO140 and MoSO160 samples with the samples produced by Sigma Aldrige[TM] gives grounds to suppose a certain similarity in their electric conduction mechanisms. The drastic change in the conduction mechanism is observed at a much higher content of the oxide/sulfoxide phase.

Note that the measurements were carried out in the dc regime and in the voltage range (0 – 10) V. The dissipated power in the samples, considering that the maximal current is no more than 1 mA, does not exceed 10 mW. Therefore, the effect of the Joule heating can be neglected. Also, it was shown that the voltage drop across the electric contacts does not exceed (1 – 10) % of the total voltage drop on the sample.

The difference between the forward and backward branches for all samples is interesting for the development of the resistive switching devices. For this purpose, we consider the dependences of the samples' resistance vs. the applied voltage in **Fig. 4**. The dependences are calculated as U/I from the I-V curves shown in **Fig. 3**.

**Figure 4(a)** demonstrates the most pronounced resistive switching effect in the samples with the NDC. The forward branch (increasing voltage) of both samples begins with a small resistance; the resistance increases sharply by an order of magnitude above ≈0.4 V; passes the maximum and saturates above 4 V or tends to saturate at the voltage higher by several times . On the backward branch (decreasing voltage), the resistance curve repeats at first the forward branch down to the maximum and then sharply increases at least by an order of magnitude more at voltages less than about 4 V and remains high down to the smallest voltage drop. A long-lasting discharge after the voltage is switched off returns the resistance to its initial state. Obtained results reveal that the $MoS_xO_y$ samples containing more than 50 % of Mo in the oxide/sulfoxide phase have metastable states characterized by a large difference in resistance, the transfer between which can be induced by applied voltage of small magnitude (several volts).

The resistive switching effect in the samples with a small content of the oxide/sulfoxide phase is much less pronounced, and the resistive switching ratio does not exceed two times. **Figure 4(b)** shows the dependences of the resistance of samples vs the applied voltage for two samples with the opposite mutual layout of the forward and backward branches. Switching between low (LR) and high (HR) resistance states is induced by applying a voltage of about 10 volts. At lower $T_s$ the switching occurs from LR to HR, and, vice versa, from HR to LR at a higher $T_s$. In the case of the MoS130 sample with the lowest $T_s$, the I-V curves, unlike other samples, are almost linear and almost without a hysteresis loop.

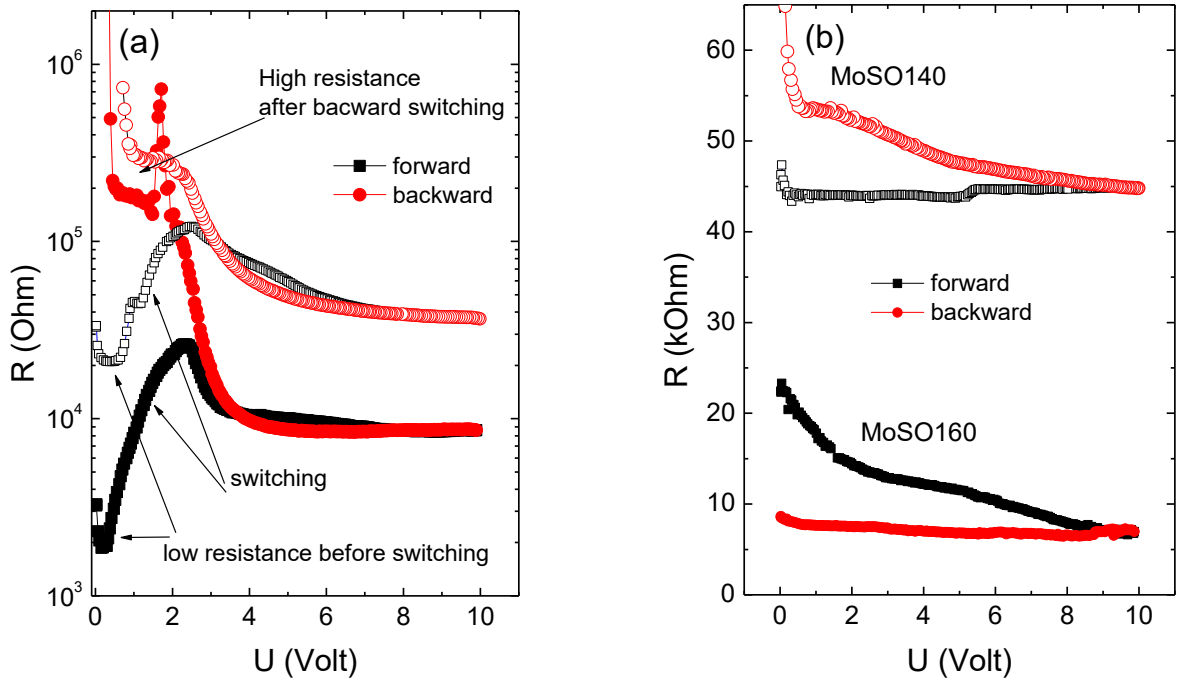

**Figure 4.** The forward and backward branches of the resistance vs. the applied voltage for the MoS150 **(a)**, MoS140 and MoS160 **(b)** samples.

The changes of the resistive switching effect in the $MoS_xO_y$ nanoflakes may be also related to the nanoparticles size. As was shown in Ref. [47], the size of nanoparticles gradually increases with an increase of $T_s$. However, this issue requires further investigations.

Thus, we reveal that the ratio of oxide/sulfoxide and sulfide/sulfate phases in the $MoS_xO_y$ nanoflake powders crucially impacts on the electric transport mechanisms in them. The increase of the Mo oxide/sulfoxide phase content above 50 % initiates appearance of the NDC mechanism with a high LR/HR ratio. This anomaly in conduction may be explained by possible changes in the energy spectrum of the electron states under the influence of oxide/sulfoxide components. The resistive switching may occur due to transitions of charge carriers between states with different mobilities or it may be due to enhanced capture or activation of charge carriers from traps. Strains, caused either by chemical structure changes or related to mechanical displacements and bending, may play an important role in both cases.

### 4.2. Accumulation of electric charge

The nanopowders of a mixed content disulfide/oxide/sulfoxide are perspective materials for supercapacitors [25]. The possibility of charge accumulation in the powders was examined by measuring temporal dependences of charging current through the sample after applying a constant voltage across the sample, and decay of voltage across the sample in the no-load regime after disconnection of the power supply circuit. Shown in **Fig. 5(a)** are the voltage decay curves across the MoSO140 sample at several initial voltages, and across the MoSO150 sample at the initial voltage drop of 2 V. Shown in **Fig. 5(b)** are corresponding to them time dependences of the charging current after switching on the constant voltage across the sample. Both measurements are made at 295 K.

As seen from **Fig. 5**, both the charge and discharge processes in the powders are long-lasting. The discharge with a decrease of the voltage by an order of magnitude has duration of several minutes. At that, one observes the two-step discharge: the first is an abrupt milliseconds long step and the second one lasts minutes. The charging process is almost the same but a little shorter and without a short initial part. The charging and discharging are described either by a simple exponential function (e.g., for the MoSO150 sample) or by a stretched exponential function (e.g., for the MoSO140 sample). Such long-lasting charging/discharging usually relates to the interfacial ion transport mechanism [17] along with the hopping conduction mechanism typical for similar materials.

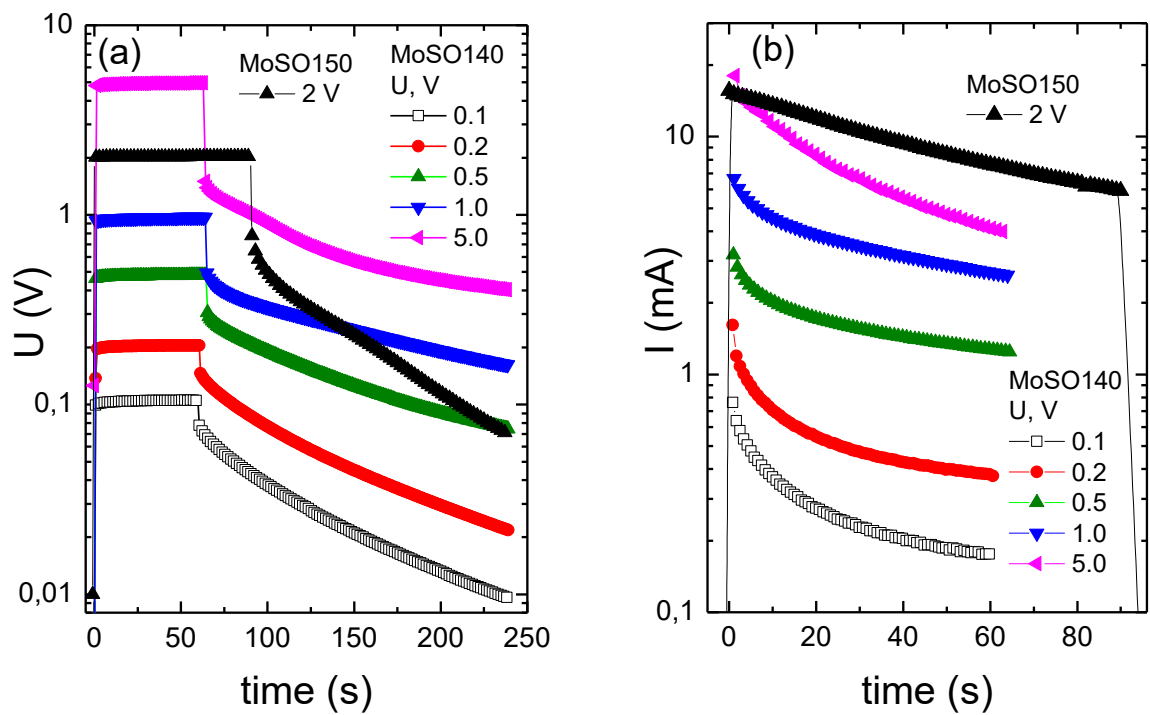

**Figure 5**. **(a)** Voltage decay across the samples in the no-load regime after disconnection of the power supply. **(b)** Charging current decay through the samples after switching on a constant voltage. Shown on the left panel are curves for the MoSO140 sample at initial voltages in the range 0.1-5 V and for the MoSO150 sample at initial voltage drop of 2 V. Measurements are performed at T=295 K.

Both mechanisms manifest characteristic changing of discharge curves with increasing temperature. **Figure 6** illustrates changes in the decay of voltage in the no-load regime for the MoSO140 sample in the temperature range from 24 to 55°C. It is seen that the decay quickly accelerates with an increase in temperature up to the total disappearance at the highest temperature. Also, it may acquire a complicated shape with increasing temperature instead of a simple exponential decay. This feature is important for applications and requires further studies aimed at improvements. The features of the charge/discharge process in the $MoS_xO_y$ nanoflakes can be additionally clarified by the frequency dependence of the capacitance of the samples.

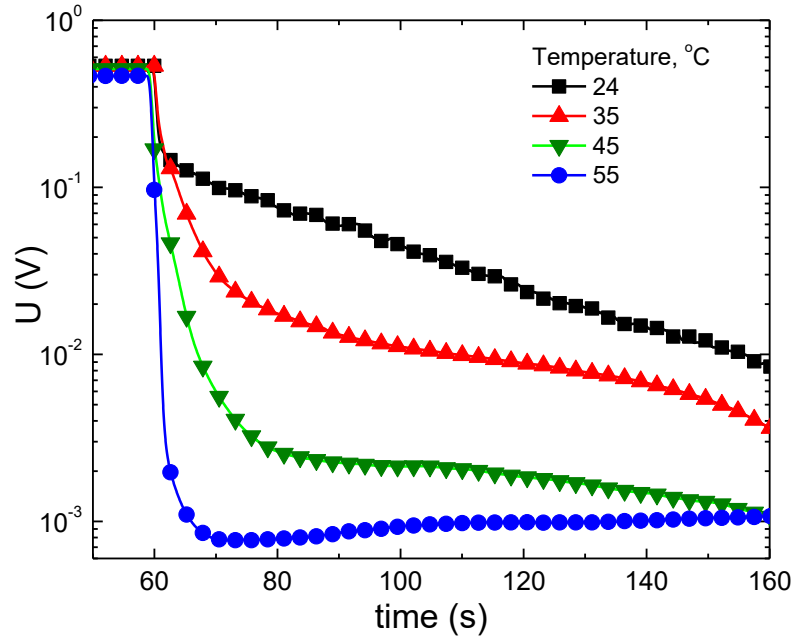

**Figure 6**. Voltage decay across the samples in the no-load regime after disconnecting the power supply at the temperature 24, 35, 45 and 55°C. The voltage magnitude before disconnection is 0.5 V. The duration of charging before the disconnection is 60 s.

### 4.3. Frequency dependences of capacitance

The frequency dependences of capacitance of all the $MoS_xO_y$ samples and the sample of pure $MoS_2$ produced by Sigma Aldrich $^{TM}$ (for comparison) are shown in **Fig. 7**. Powder samples were pressed into tablets by uniaxial pressure of 2.5 MPa. All samples have a disk shape with the same diameter of 4 mm and thickness 150 μm. The curves in **Fig. 7** show the behavior of the dielectric permittivity.

The capacitance and, respectively the dielectric permittivity, decrease approximately proportional to the reciprocal frequency for all samples. The dielectric permittivity differs among samples with different phase composition by four orders of magnitude. At that, one may distinguish between two groups with high and low capacitance and dielectric permittivity. The $MoS_xO_y$ samples sintered at $T_s$ = 140, 150 and 180°C have higher capacitance and permittivity; meanwhile the $MoS_xO_y$ samples sintered at $T_s$ = 160°C have the permittivity several order of magnitude less. The permittivity of the sample prepared from the powder of $MoS_2$ produced by Sigma Aldrich $^{TM}$ demonstrates the behavior coinciding with the low capacitance group.

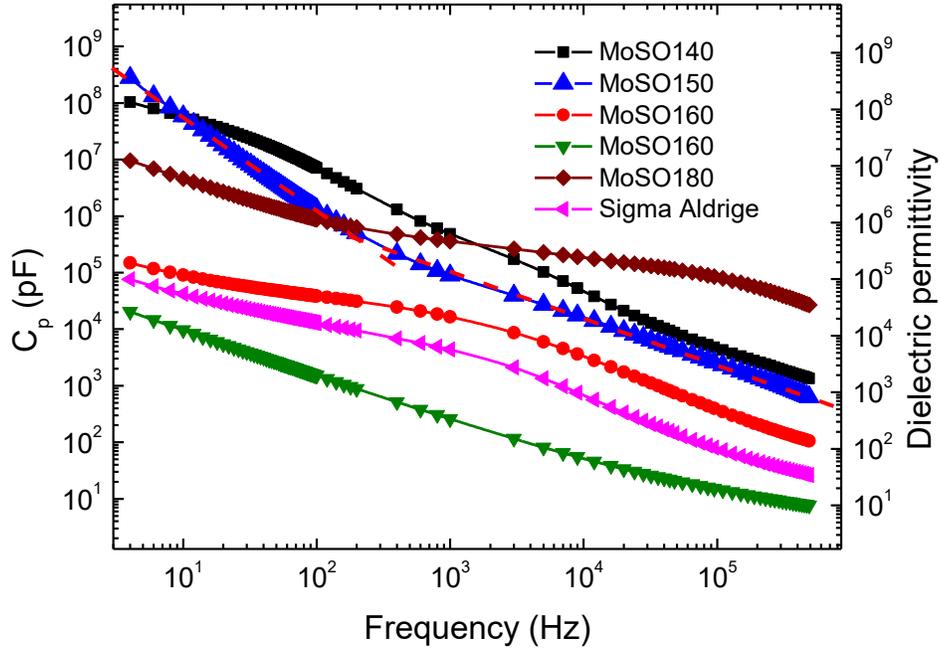

**Figure 7**. Frequency dependences of the capacitance (left scale) and dielectric permittivity (right scale) for the studied $MoS_xO_y$ samples and the $MoS_2$ sample produced by Sigma Aldrich $^{TM}$.

Special attention attracts the MoSO150 sample. It has two clearly seen parts in the frequency dependence of capacitance. It also has the highest capacitance (and thus, permittivity) at the lowest frequency. First of all, note the giant permittivity observed at the lowest frequency (4 Hz) which is about $3 \cdot 10^8$. The dependence has two clearly distinguished parts $\varepsilon \sim f^{-1.64}$ and $\varepsilon \sim f^{-0.77}$, which is illustrated by two dash lines at low and higher frequency ranges.

The frequency dependences of capacitance for all samples studied have no horizontal parts at low frequencies. Such unusual dependences were observed in the nanopower samples of some ferroelectric and/or spinel materials (see, e.g., Ref. [48]). However, such frequency dependency was unknown earlier for $MoS_2$ and $MoS_xO_y$ powders.

### 4.4. Thermo-emf

To clarify the character of charge carriers participating in the conduction of the $MoS_xO_y$ samples we performed also measurements of the thermo-emf. Also, for comparison, the thermo-emf was measured for the $MoS_2$ sample prepared from the Sigma Aldrich $^{TM}$ powder. The results are shown in **Fig. 8.** It is seen that thermo-emf of $MoS_xO_y$ samples differs by the sign from that of the $MoS_2$ sample. While the thermo-emf has an entirely negative sign corresponding to the hole-type conduction in the MoS$_2$ powder, the MoS$_x$O$_y$ samples manifest predominantly the electron-type conduction, which is possibly mixed with the hole-type conduction under heating or at small ΔT, measured in regard to 295K, or in a strong electric field.

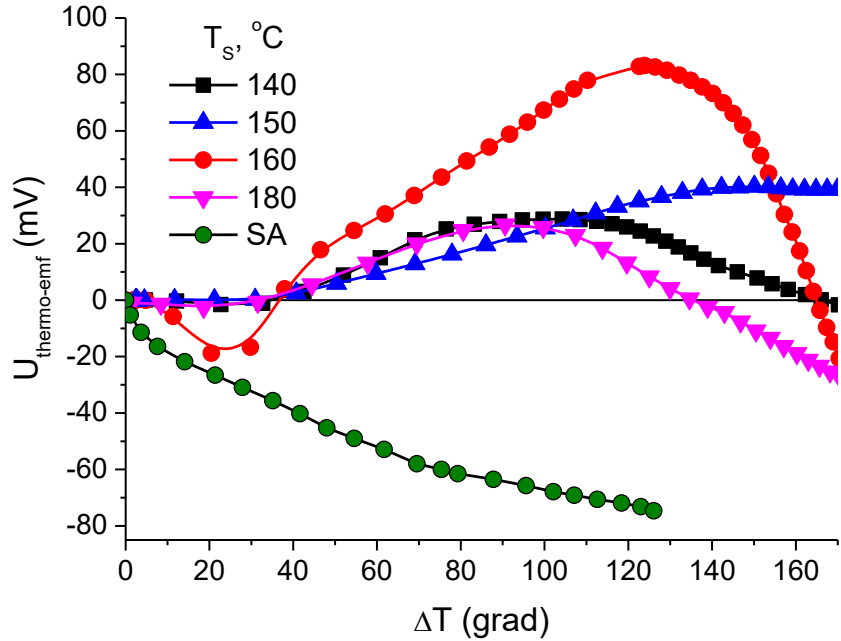

**Figure 8.** Dependence of the thermo-emf voltage vs ΔT for different studied MoS$_x$O$_y$ samples and for the MoS$_2$ sample (SA) produced by Sigma Aldrich ™. ΔT is measured in regard to 295K.

In a small initial range of ΔT the thermo-emf voltage curves measured for MoS$_x$O$_y$ samples may have a very small magnitude of thermo-emf with the negative sign, except for the sample with Ts=160 °C, which has a pronounced negative thermo-emf. The sample with Ts=150 °C has a very small positive thermo-emf in this interval of ΔT. With further increase in ΔT the thermo-emf has a positive sign and grows with different slope for all MoS$_x$O$_y$ samples. At ΔT =100 – 130 degree the curves for Ts=140, 160 and 180°C change their slope to negative and tend to the region of the negative thermo-emf. Such behavior gives evidence of the mixed character of electric conductivity, both electron- and hole-type components. At that the ratio between these components depends on the temperature and stress in the nanoflakes agglomerates. It is confirmed, e.g., by the fact that one demands a long time to restore initial characteristics after heating. At the same time, the reproducibility and stability of results in the initial state after compressing tablets (at the same uniaxial pressure of 2.5 MPa) before the measurements was checked during several hours.

Theory of the thermo-emf effect is well developed in semiconducting materials (see e.g., Ref. [49]). In the case of semiconductors, the differential thermo-emf has a magnitude of the order of 0.1-1 mV/K, which agrees with that observed both in the samples of $MoS_2$ powders produced by Sigma Aldrich [TM] and in the $MoS_xO_y$ samples under consideration. In the case of mixed kind with participation both of electron and hole conduction types the differential thermo-emf acquires a complicated behavior and obeys the formula:

$$\alpha = \frac{\alpha_n \sigma_n + \alpha_p \sigma_p}{\sigma_n + \sigma_p}, \quad (1)$$

where $\sigma_n = e\mu_n n$ and $\sigma_p = e\mu_p p$ are the electron and hole contributions to electric conductivity. In the case of the sulfo/oxide molybdenum powder samples with a strong temperature dependence of resistivity one may suppose participation of the hopping kind conduction along with the band one. Therefore, the mobilities $\mu_n$ and $\mu_p$ should be considered rather as phenomenological parameters. The observed dependences of the thermo-emf on $\Delta T$ evidence of non-monotonous change of the mobilities and concentrations of the charge carriers participating in conduction.

**4.5. Summary of electric conduction studies**

To summarize the studies of electric conduction mechanisms, we would like to underline the following. Due to the absence of data describing the electron states energy spectrum in the nanoflakes with such oxygen-containing forms, it is difficult to give completely comprehensive explanation of some peculiarities of the conduction mechanism and its difference in connection with the samples preparation conditions. It is not clear why the $MoS_xO_y$ samples, synthesized at $T_s = 150$ °C, have the highest content of oxide/sulfoxide component compared to those synthesized at lower and higher temperatures, and what changes in the energy spectrum of the electron states appear due to this. Since the effect is reproducible in different samples synthesized at $T_s = 150$°C and does not degrade with time, this question requires further studies.

The samples with a smaller content (≈25 %) of Mo in oxide/sulfoxide compounds have loop-like I-V curves. They look very similar to the I-V curves of the chemically pure $MoS_2$ produced by Sigma Aldrich [TM] [17]. At the same time, they have a different kind of resistive switching, either LR/HR or HR/LH, depending on the synthesis temperature $T_s$. Such changes presumably may be related to the mechanisms of the charge transfer between metastable states or to the appearance of highly conductive filaments. The samples with the N-shaped I-V curves are promising for memristor applications [50]. In particular, except for the high LR/HR ratio, they provide a possibility to realize a three-level memristor.

Also, the $MoS_xO_y$ samples have perspectives for application in charge accumulation devices, because they demonstrated very long-lasting charging/discharging. The samples synthesized at relatively low $T_s$ have very large specific capacity. The two-step behavior is inherent for their discharging. The first step of an abrupt jumping decrease may be related to the fast polarization processes unrelated to the redistribution of the ion space charge. The second one may be related to ion transport in the interfaces between nanoflakes and their aggregates [17]. The second one is longer in the samples without NDC.

## 5. THEORETICAL MODELLING

Modern theory of resistive switching originates from Strukov et al. [51, 52, 53], who demonstrated that memristive behavior can be inherent to thin semiconductor films, at that the memory resistance depends on the thickness ratio of the doped and pure regions of the semiconductor.

Note, that the flexo-chemical strains can be the reason of resistive switching observed in thin semiconducting films with a mixed ionic-electronic conductivity [54], and the bending effects can influence very strongly the conductivity of the LD TMD [36]. Since we observed I-V curves, which belong to memristor type [55, 56], the model describing the memristive switching should be used.

Though one can expect pronounced resistive switching effects and strong correlation between the nonlinear current-voltage and strain-voltage response of the suppressed TMD nanoflakes, the latter was not studied theoretically. Intentions to fill the gap in the knowledge motivated us to perform self-consistent modelling of nonlinear electric transport and electromechanical response of the oxidated TMD nanoflakes allowing for steric effects of mobile defects (i.e., aggregation of ions or vacancies), flexoelectricity and flexo-chemical coupling with the Vegard strains and/or strains appeared (or restored) during the nanoflakes formation.

To apply the memristive switching models to describe the experimentally observed I-V curves one should use some hypothesis of the possible origin and physical nature of the mobile charges in the studied nanoflakes.

The strong bending of the oxidized free-standing nanoflake may occur due to the flexo-electric field effect, and the effect can also change the bending degree up to the appearance of the symmetry lowering in the nanoflake. Notably, that the effect magnitude can be critically sensitive to the concentration of elastic defects, which (in their turn) can depend very strongly and non-monotonously on the synthesis conditions (especially when we speak about oxygen states).

According to the model of sliding flexo-ferroelectricity [41], the charge $Q_{flexo}$ accumulated by the nanoflake due to the flexoelectric coupling can be estimated from the following relation:

$$Q_{flexo} \approx \iiint_0^V div(\vec{P}_{flexo})dV \approx \iiint_0^V f_{ijkl}\frac{\partial^2 u_{kl}^{flexo}}{\partial x_i \partial x_j}dV \approx \iiint_0^V f_{ijkl}\left(q_{klmn}\frac{\partial^2}{\partial x_i \partial x_j}(E_k^{loc}E_l^{loc}) + z_{klmnop}\frac{\partial^2}{\partial x_i \partial x_j}(E_k^{loc}E_l^{loc}E_o^{loc}E_p^{loc}) + z'_{klmnoprs}\frac{\partial^2}{\partial x_i \partial x_j}(E_k^{loc}E_l^{loc}E_o^{loc}E_p^{loc}E_r^{loc}E_s^{loc}) + \cdots\right)dV, \quad (2)$$

where $V$ is the nanoflake volume, $\vec{P}_{flexo}$ is the flexoelectric polarization, $f_{ijkl}$ is the flexoelectric tensor, $u_{kl}^{flexo}$ is the flexoelectric deformation, $q_{klmn}$ is the second-order electrostriction tensor, $z_{klmnop}$ is the fourth-order electrostriction tensor, etc. Notably, strong bending requires the consideration of the higher-order electrostriction couplings. $E_k^{loc}$ is the component of the local electric field induced by the inhomogeneous (but regular) external field and randomized flexo-chemical field, i.e., $E_k^{loc}(\vec{r},t) = E_k^{ext}(\vec{r},t) + \delta E_k^{f-c}(\vec{r},t)$. Hereinafter, we consider the case when the flexo-chemical strain is linearly proportional to the distributed concentration of the random elastic defects $\delta C(\vec{r})$ (the Vegard law for chemical strains [57]). The statistic average $\langle\langle\frac{\partial^2}{\partial x_i \partial x_j}\delta C\rangle\rangle = 0$ (as well as for all odd powers).

The calculations of the local field are very complex, since the field depends in a self-consistent way on the flake bending, defects concentration and distribution, as well as on the electric state of surrounding nanoflakes and their concentration. Recently, the thermodynamic Landau-Cahn-Hilliard phase field model has been used for the self-consistent simulations of memristive thin film morphology and current-voltage hysteresis [58]. Without any restrictions on conducting filaments geometry (if any), the model [58] predicts that the conducting filaments can emerge on thermodynamic paths, which are energetically favorable due to structural and chemical variations occurring during their preparation.

The application of the Landau-Cahn-Hilliard approach [58] to the $MoS_xO_y$ nanoflakes considering the inhomogeneous flexo-chemical field leads to inclusion of the local field $E_k^{loc}(\vec{r},t)$ in the Cahn-Hilliard functional (see Eq.(5) in Ref. [58]). Phenomenological Cahn-Hilliard equations can be obtained after the statistical averaging.

Since the regular part of the local field is induced by the application of the electric voltage $U$, which can change the flake shape, it is reasonable to assume that its amplitude is proportional to the applied voltage, $E_k^{loc} \sim U(1 + \delta C)$ at small $U$, and saturates under the voltage increase. Assuming zero statistic average $\langle\langle \delta E_k^{f-c}(\vec{r},t)\rangle\rangle = 0$, and the nonzero mean-square statistic average $\langle\langle \delta C \frac{\partial^2}{\partial x_i \partial x_j} \delta C \rangle\rangle \neq 0$, the $Q_{flexo}$ can be expanded in series over the even powers of the applied voltage:

$$Q_{flexo} \cong \frac{Q_1 U^2 + Q_2 U^4 + Q_3 U^6}{1 + D_3 U^6}, \qquad (3)$$

where the coefficients $Q_i$ are proportional to the corresponding integrals in Eq.(2), e.g., $Q_1 = \frac{1}{U^2} \iiint_0^V f_{ijkl} q_{klmn} \langle\langle \frac{\partial^2}{\partial x_i \partial x_j} E_k^{loc} E_l^{loc} \rangle\rangle dV$, etc., and the constant $D_3$ originates from the local field saturation at high voltages.

Equation (3) describes phenomenologically the nonlinear static dependence of the accumulated charge, which is one possible contribution to the electric charge-discharge process. Other charge accumulation processes, leakage and discharge mechanisms may be responsible for the hysteresis-like differences of the forward and backward branches of the I-V curves (see e.g., Ref.[59]). Following Ref. [59], we can assume that the space charges can be trapped by the sites of different shapes and sizes corresponding to interfaces and intersections of nanoflakes in the compressed powder. The charge dynamics is conditioned by the charge trapping in one site and release from the site due to the leakage effect, as well as by the trapping in another site and eventual escape from the capacitor after a huge amount of trapping and release, further associated with the hopping conduction mechanism. A negligibly small number of charges can enter and exit the capacitor with TMD nanoflakes without undergoing trapping-and-release steps.

Considering the hopping and leakage effects, as well as the sliding flexo-ferroelectric and flexo-chemical contribution to the charge accumulation, the electric current density $j$ can be estimated as:

$$\vec{j} = \varepsilon_0 \frac{d\vec{E}}{dt} + \frac{d\vec{P}_{flexo}}{dt} + \vec{j}_{hopping} + \vec{j}_{leakage}. \qquad (4a)$$

Here $\varepsilon_0$ is the universal dielectric constant. Corresponding electric current is [59]:

$$I = C_0 \frac{dU}{dt} + \frac{dQ_{flexo}}{dt} + \left(G_0 + \frac{e\mu}{L^2}\delta n\right)U. \quad (4b)$$

Here we used that $E = U/L$, $C = \varepsilon_0 \varepsilon S/L$ is the effective capacitance ($L$ is the width and $S$ is the surface area, $V = SL$). The sum $G_0 + \frac{e\mu}{L^2}\delta n$ is the residual leaking conductance and the hopping mobility contributions. It maybe $G_0 \gg \frac{e\mu}{L^2}\delta n$, where $\mu$ is the mobility of nonequilibrium carriers with the charge density $e\delta n$.

Using Eq.(3) and (4b), the phenomenological relation for the I-V curves fitting acquires the form:

$$I(U) \approx \frac{U}{R_0} + \frac{dU}{dt}\left(C_0 + \frac{2Q_1 U + 4Q_2 U^3 + 6Q_3 U^5}{1+D_3 U^6}\right) + \frac{1}{1+D_3 U^6}\left(\frac{dQ_1}{dt}U^2 + \frac{dQ_2}{dt}U^4 + \frac{dQ_3}{dt}U^6\right), \quad (5a)$$

where $R_0$ is the resistance in the linear approximation (without consideration of the sliding flexo-ferroelectricity and elastic defects), and so $\frac{1}{R_0} \cong G_0 + \frac{e\mu}{L^2}\delta n$. The nonlinear terms originate from the flexo-chemical strains. Hereinafter the phenomenological coefficients $Q_i$ and the saturation rate $D_3$ are treated as the fitting parameters for I-V curves, which signs and values depend on the synthesis conditions. From Eq.(5a), the resistivity can be introduced as:

$$R(U) = \frac{U}{I(U)}. \quad (5b)$$

If the electric discharge time $\tau$ is much longer than the voltage step delay $\tau_V$, we can use the approximate equalities $\frac{dU}{dt} \approx \frac{U}{\tau_V}$ and $\frac{dQ_i}{dt} \approx \frac{Q_i}{\tau}$ and the strong inequality $\frac{1}{\tau} \ll \frac{1}{\tau_V}$ in Eq.(5a). Indeed, the voltage step delay is 2 seconds, and the discharge time is much more than 20 seconds (see **Fig. 9**). The circumstance allows to approximate Eqs.(5) in the quasi-static limit as:

$$I(U) \approx U\left(\frac{1}{R_0} + \frac{1}{\tau_V}\left[C_0 + \frac{2Q_1 U + 4Q_2 U^3 + 6Q_3 U^5}{1+D_3 U^6}\right]\right), \quad (6a)$$

$$R(U) = \frac{R_0}{1+\frac{R_0}{\tau_V}\left[C_0 + \frac{2Q_1 U + 4Q_2 U^3 + 6Q_3 U^5}{1+D_3 U^6}\right]}. \quad (6b)$$

Equations (6) describe phenomenologically the nonlinear static dependence of the accumulated charge, but they cannot explain the differences between the forward and backward branches of the I-V and R-V curves observed experimentally (see **Figs. 2** and **3**). To do this a phase delay between the ohmic resistance and capacitance should be considered. The delay leads to a more cumbersome equation for complex amplitudes of the current ($\tilde{I}$) and voltage ($\tilde{U}$):

$$\tilde{I}(U) \approx \frac{\tilde{U}}{R_0}\left(1 + \frac{C_1 U + C_2 U^3 + C_3 U^5}{1+D_3 U^6} + i\omega R_0 C_0\left[1 + \frac{D_1 U + D_2 U^3 + D_3 U^5}{1+D_3 U^6}\right]\right), \quad (7)$$

The fitting parameters $C_i$ are related with the parameters $D_i$ as $D_1 = 2\frac{C_1 C_0}{\tau}$, $D_2 = 4\frac{C_2 C_0}{\tau}$ and $D_3 = 6\frac{C_3 C_0}{\tau}$. Note that $\tau$ can be a fitting parameter.

The I-V curves and resistivity calculated from Eqs.(7) for the curved TMD nanoflakes at increasing magnitude of the flexo-chemical strain are shown in **Fig. 9**. Red and blue curves correspond to the direct (increase) and inverse (decrease) direction of the voltage cycling. It is seen that the shape of the I-V curves, their slope, current and voltage scales resemble the experimentally measured I-V curves shown in **Fig. 2**. The increase in the flexo-chemical strains leads to the transition of the quasi-linear I-V curves (**Fig. 9(a)**) to the curves with the NDC region in the actual voltage range (**Fig. 5b**). The voltage dependences of the corresponding resistances contain the regions of NDC at least for the direct run of the voltage sweep (see **Figs. 5(c-d)**). To summarize, the proposed model may explain the observed polar and electronic properties of the oxidized TDM nanoflakes.

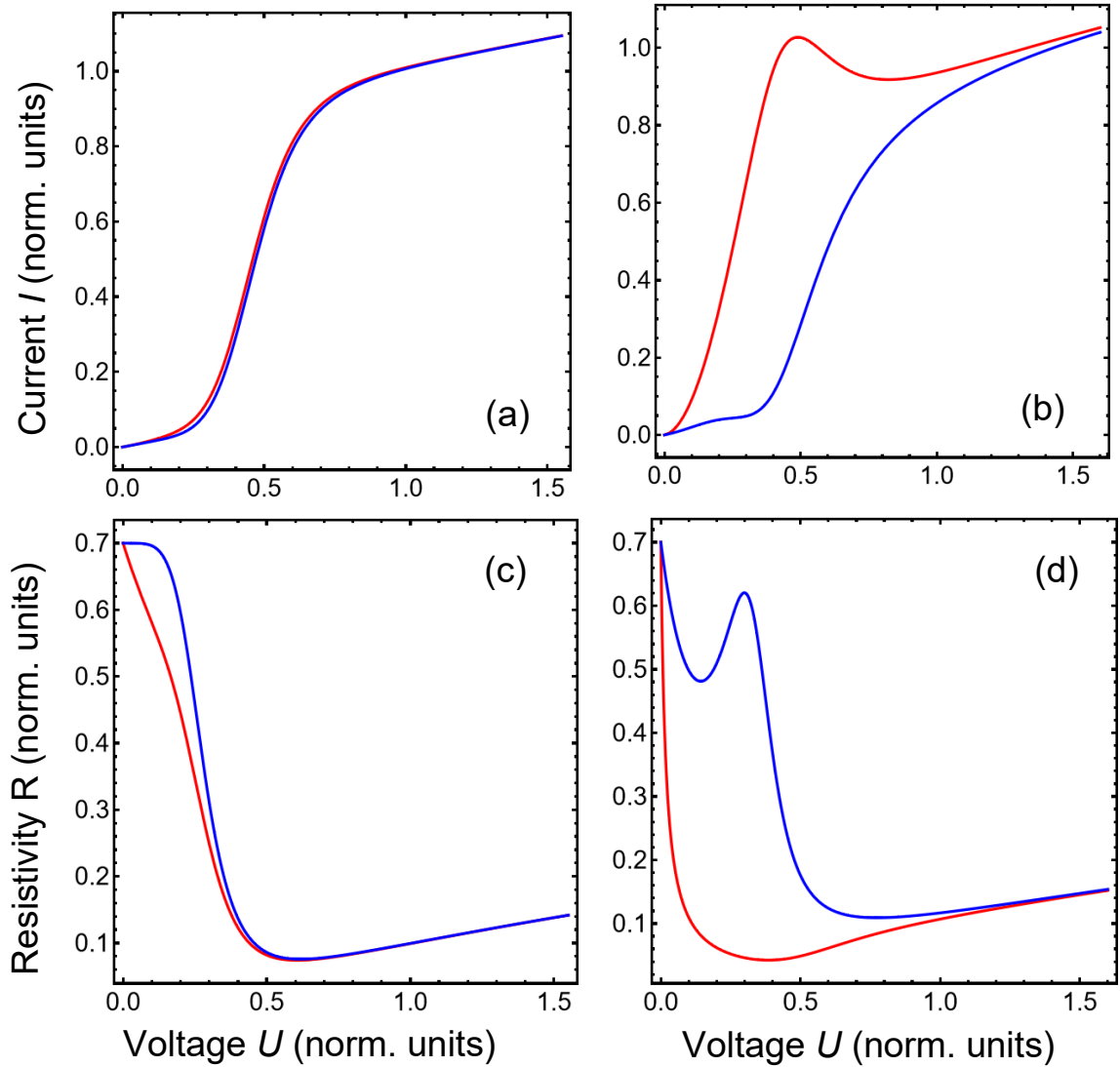

**Figure 9.** I-V curves **(a-b)** and resistivity **(c-d)** calculated for the oxidized TMD nanoflakes at increasing magnitude of the flexo-chemical strain, which is small for the plots **(a, c)**, and large for the plots **(b, d)**. Red and blue curves correspond to the direct (increase) and inverse (decrease) direction of the voltage sweep. The plots are calculated in normalized units. Fitting parameters are listed in **Table 3**.

**Table 3.** Fitting parameters (in dimensionless units)

| Fitting parameter | Small flexo-chemical strain | Large flexo-chemical strain |
|---|---|---|
| $R_0$ | 70 | 70 |

| $C_1$ | 0.2 (forward) – 0 (backward) | 5.5 (forward) – 0.55 (backward) |
| $C_2$ | 0 | -0.1 |
| $C_3$ | 0.0055 | 0.0055 |
| $D_3$ | 0.00009 | 0.00009 |

## 6. SUMMARY

We analyzed the electric transport mechanisms in the powders of the molybdenum-disulfide-oxide nanoflowers consisting of self-assembled 10-20 nm thin nanoflakes prepared by the hydrothermal reaction of $(NH_4)_6Mo_7O_{24}$ with thiourea in aqueous solution and synthesized at different temperatures from 130°C to 180°C. According to the XPS, EDS and Raman spectroscopy studies presented in Ref. [47], a significant amount of the molybdenum oxides and sulfoxides in the presumably molybdenum disulfide nanoflakes allows to consider the "effective" formulae $MoS_xO_y$ for their chemical composition.

The studied $MoS_xO_y$ nanoflakes reveal the electric transport features interesting for material science and promising for applications. The chemical composition and microstructure of the nanoflakes, which have been determined by the temperature of synthesis and the efficiency of oxidation in ambient conditions, are shown to be responsible for the changes in the electric conduction mechanism. The current-voltage characteristics of $MoS_xO_y$ nanoflakes possess hysteresis behavior with a loop between the forward (increasing voltage) and backward (decreasing voltage) branches. The parameters of hysteresis loop are determined by the degree of oxidation corresponding to the content of oxygen containing components.

The samples with low and high ratio of disulfide and oxide/sulfoxide phases have significantly different behavior of the I-V curves. So, the samples characterized by a high (>50 %) content of the oxide/sulfoxide phases possess a strongly pronounced negative differential conductivity (N-shape) part in the I-V curves and a much higher LR/HR ratio additionally to the hysteresis loop behavior. Important result is that the $MoS_xO_y$ nanoflakes manifest very long living deep charging and discharging after the voltage across the sample switching "on" and "off".

It is well-known that the presence of oxygen non-stoichiometry and bending can create strong flexo-chemical strains in the individual nanoflakes in the pressed state of their dense powder. Taking this into account, we modified the thermodynamical Landau-Cahn-Hilliard approach [58] for the description of resistive switching and electro-transport properties in thin films by considering flexo-chemical strains induced by the chemical/compositional strains and flexoelectric effect in the pressed $MoS_xO_y$ nanoflakes. In result we obtained the phenomenological expressions, which describe the form of the observed I-V curves.

The features of memristive switching and charge accumulation in the $MoS_xO_y$ nanoflakes, revealed experimentally and explained theoretically in the framework of the modified Landau-Cahn-Hilliard approach, look promising for applications in memristors and high-performance supercapacitors.

**Supplementary Material.** Supporting Information containing data of XPS (**Appendix A1**), SEM and EDX (**Appendix A2**), Raman spectroscopy and TEM (**Appendix A3**) are given in Supplementary Materials.

**Author contributions.** O.S.P., V.V.V. and V.N.P. conceived, performed and analyzed results of the electrophysical experiments. A.V.T. and S.V.K. prepared the samples and characterized them by SEM and EDS. A.S.N. and V.I.P. performed Raman studies. A.S.N., M.V.O. and G.I.D. analyzed the spectra. G.I.D. and O.B. performed TEM and analyzed obtained results. A.S.T. performed XRD analysis. T.S. and B.M.R. performed XPS measurements. A.N.M. proposed the model for interpretation of the experimental results. V.V.V. and A.N.M. wrote the manuscript draft, and all authors participated in its improvement.

**Acknowledgments.** Synthesis of samples (A.V.T. and S.V.K.) was supported by the National Academy of sciences, project "Development of composites based on sulfides of metals of the 6 and 7 group and porous carriers for liquid-phase selective catalytic hydrogenation of halogen-containing heterocyclic compounds. Structural study (G.I.D) was supported by Joensuu Foundation Project Decision number 359463. Electrophysical measurements are supported by the Target Program of the National Academy of Sciences of Ukraine, Project No. 5.8/25-П "Energy-saving and environmentally friendly nanoscale ferroics for the development of sensorics, nanoelectronics and spintronics" (O.S.P., V.N.P., and V.V.V). The theoretical work of A.N.M. is supported by the DOE Software Project on "Computational Mesoscale Science and Open Software for Quantum Materials", under Award Number DE-SC0020145 as a part of the Computational Materials Sciences Program of US Department of Energy, Office of Science, Basic Energy Sciences; and by the Horizon Europe Framework Programme (HORIZON-TMA-MSCA-SE), project № 101131229, Piezoelectricity in 2D-materials: materials, modeling, and applications (PIEZO 2D). Analytical results, presented in this work, are visualized in Mathematica 14.2 [60].

[59] O. Lipan, F. Hartmann, S. Höfling, and V. Lopez-Richard. Unveiling Ferroelectric-Like Behavior in Leaky Dielectrics: A Microscopic Model for Polarization Dynamics and Hysteresis Inversion. *arXiv preprint*, arXiv:2410.16084 (2024). https://doi.org/10.48550/arXiv.2410.16084.

[60] The Mathematica (https://www.wolfram.com/mathematica) notebook, which contain the codes, is available per reasonable request.
33

# Supplementary Materials to

# "The impact of morphological structure and flexo-chemical strains on the electric transport mechanisms in the molybdenum-disulfide-oxide nanoflakes"

O. S. Pylypchuk[1], V. V. Vainberg[1a], V. N. Poroshin[1], A. V. Terebilenko[2], A. S. Nikolenko[3], V. I. Popenko[3], A. S. Tolochko[1], M. V. Olenchuk[1], O. Bezkrovnyi[4], G. I. Dovbeshko[1], T. Sabov[3], B. M. Romanyuk[3], S. V. Kolotilov[2b] and A. N. Morozovska[1c]

[1] Institute of Physics, National Academy of Sciences of Ukraine, 46, pr. Nauky, 03028 Kyiv, Ukraine

[2] L.V. Pisarzhevskii Institute of Physical Chemistry, National Academy of Sciences of Ukraine, 31, pr. Nauky, 03028 Kyiv, Ukraine

[3] Lashkarev Institute of Semiconductor Physics, National Academy of Sciences of Ukraine, 41, pr. Nauky, 03028 Kyiv, Ukraine

[4] W. Trzebiatowski Institute of Low Temperature and Structure Research, Polish Academy of Sciences, 50-422 Wroclaw, Poland

## Appendix A

**A1. XPS measurements**

The survey spectra for all samples are presented in **Fig. S1** and **S2**. The analysis of the Mo 3d and S 2p regions indicated the presence of oxygen, which can be attributed to molybdenum oxide ($MoO_3$) (see Ref. [1]). Also, the XPS spectrum shows the presence of some quantity of carbon. Its content varies randomly and does not depend on the Ts. It can originate from absorbed thiourea or carbonates (the product of hydrolysis of thiourea).

[a] corresponding author, e-mail: viktor.vainberg@gmail.com
[b] corresponding author, e-mail: s.v.kolotilov@gmail.com
[c] corresponding author, e-mail: anna.n.morozovska@gmail.com (submitting author)

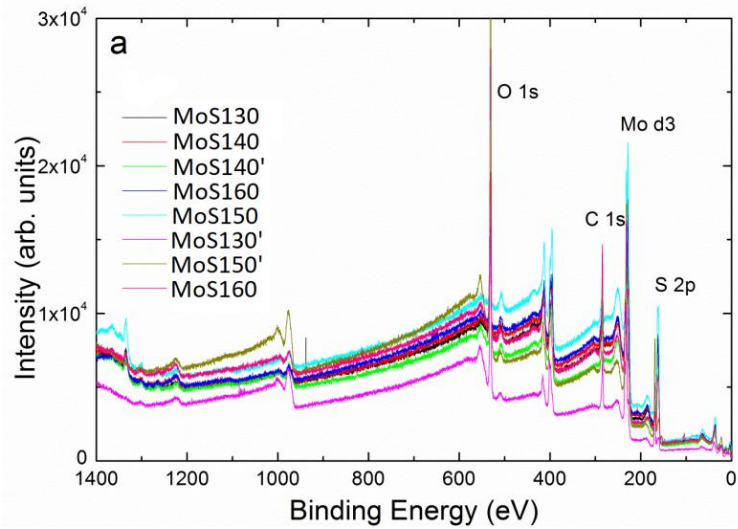

**Figure S1.** XPS spectra the MoSO nanoflowers formed at temperatures $T_S$ from 130 ºC to 160ºC. The analysis of the MoSO150 spectrum reveals the presence of molybdenum disulfide along with the large amount of oxidized Mo-containing species as well as oxidized sulfur.

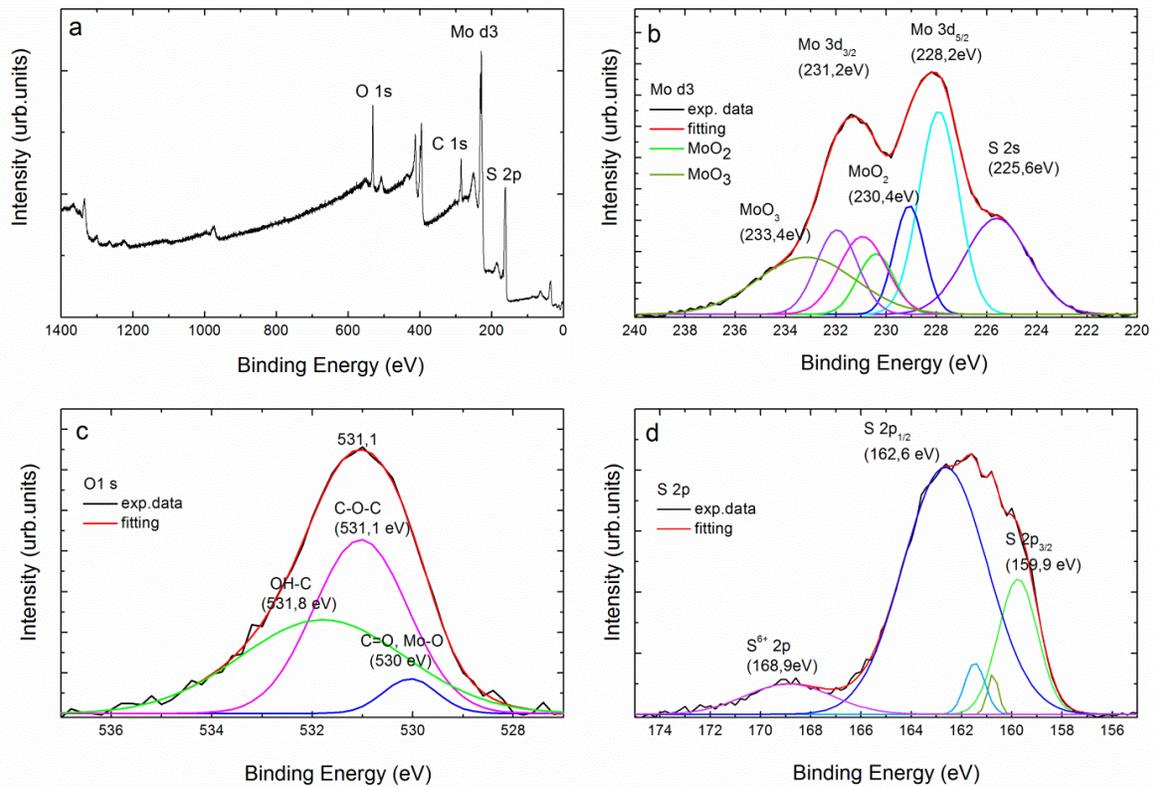

**Figure S2. (a)** Typical survey XPS spectrum of MoSO150. The XPS spectrum in theregions of Mo 3d **(b)**, O 1s **(c)** and S 2p **(d)**. Red curves are the fitting of the experimental (black) curves. The curves of other colors show separate peaks decomposed from the red curves into the Voight peaks.

## A2. Electron microscopy and EDS

The SEM images of the MoSO140, MoSO150 and MoSO160 samples are shown in **Figs. S3 (a)**, **(b)** and **(c)**, respectively. As can be seen, the particles in all samples display generally consistent morphological features and tendency to aggregation. Some distinctive features in **Fig. 3(b)** for MoSO150 may be highlighted as compared to the other two materials. The image looks denser and "brighter" (even with metallic shine), and the particles look smaller. The left and right images are "darker" and have larger blocks of particles, relatively better separated. These features may be associated with significantly higher content of oxidized compounds.

The EDX spectra of the same samples are shown in **Fig. S3(d)**, **(e)** and **(f)**, respectively. All spectra contain the peak at 2.4 eV attributed to the molybdenum (Mo). However, the EDX spectrum of the MoSO150 nanoflakes reveals a distinctive feature as compared to other two. It contains smaller content of oxygen (the data are presented in **Table S1**).

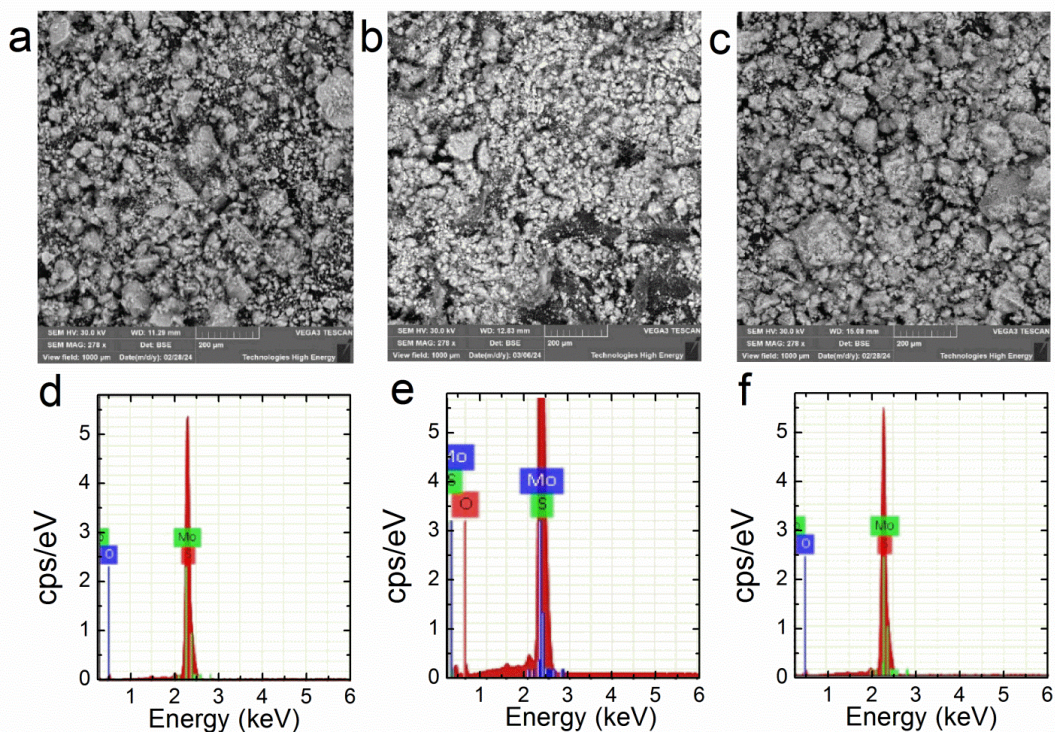

**Figure S3**. The survey SEM images of the MoSO140 **(a)**, MoSO150 **(b)**, MoSO160 **(c)** and the EDX spectra of the MoSO140 **(d)**, MoSO150 **(e)** and MoSO160 **(f)**, respectively. The scale is 200 µm.

The content of O, S and Mo in the samples MoSO140, MoSO150 and MoSO160, calculated from t EDX spectra obtained in 3 points, is listed in **Table T1.**

**Table T1.** Element content in the three points of the $MoS_xO_y$ nanoflake powders MoSO140, MoSO150 and MSO160.

| Sample | Element | Mass % | At % |
|---|---|---|---|
| MoS140 | O | 20.14; 18.43; 17.23 | 44.14; 41.42; 39.85 |
|  | S | 36.66; 37.52; 36.71 | 40.08; 42.07; 42.38 |
|  | Mo | 43.19; 44.05; 46.06 | 14.78; 16.51; 17.77 |
| MoS150 | O | 16.25; 14.45; 13.94 | 38.30; 34.31; 33.40 |
|  | S | 36.75; 40.34; 40.51 | 43.22; 47.79; 48.41 |
|  | Mo | 47.00; 45.21; 45.54 | 18.48; 17.90; 18.19 |
| MoS160 | O | 20.31; 19.88; 21.29 | 45.78; 45.51; 46.73 |
|  | S | 32.41; 31.42; 33.57 | 36.45; 35.89; 36.76 |
|  | Mo | 47.28; 48.71; 45.13 | 17.77; 18.60; 16.52 |

Typical TEM images of the nanoflakes MoS150 are shown in **Fig. S4**. The profiles of the sharp and very contrast edges correspond to the small thickness (about 10 nm or less) and strong bending of the individual nanoflakes. Small sizes of the particles favor their oxidation in ambient conditions.

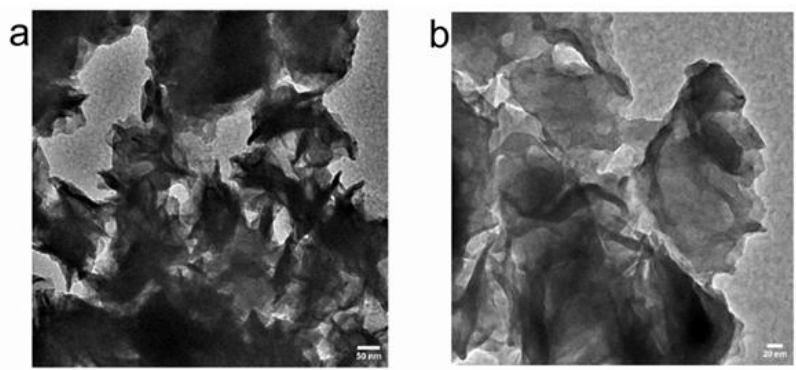

**Figure S4.** Typical TEM images of MoSO150 with lower (a) and higher (b) magnifications.

### A3. Raman spectroscopy

The Raman spectra of the pressed MoSO140, MoSO150 and MoSO160 samples, being the most representative for understanding electro-transport features, are shown in **Fig. S5** with their analysis in **Table T2**. The analysis is consistent with Refs. [2-6].

To summarize, the Raman spectra confirm the presence of molybdenum oxide and sulfoxide groups on the surface of all $MoS_xO_y$ nanoflakes (as well as their self-assembling in nanoflowers) in high amount, as follows from the relative intensities of corresponding peaks. The fraction of the molybdenum oxides, sulfoxides and/or other possible oxygen-containing compounds non-monotonically depends on the temperature of synthesis showing distinctive features for the MoS150 sample.

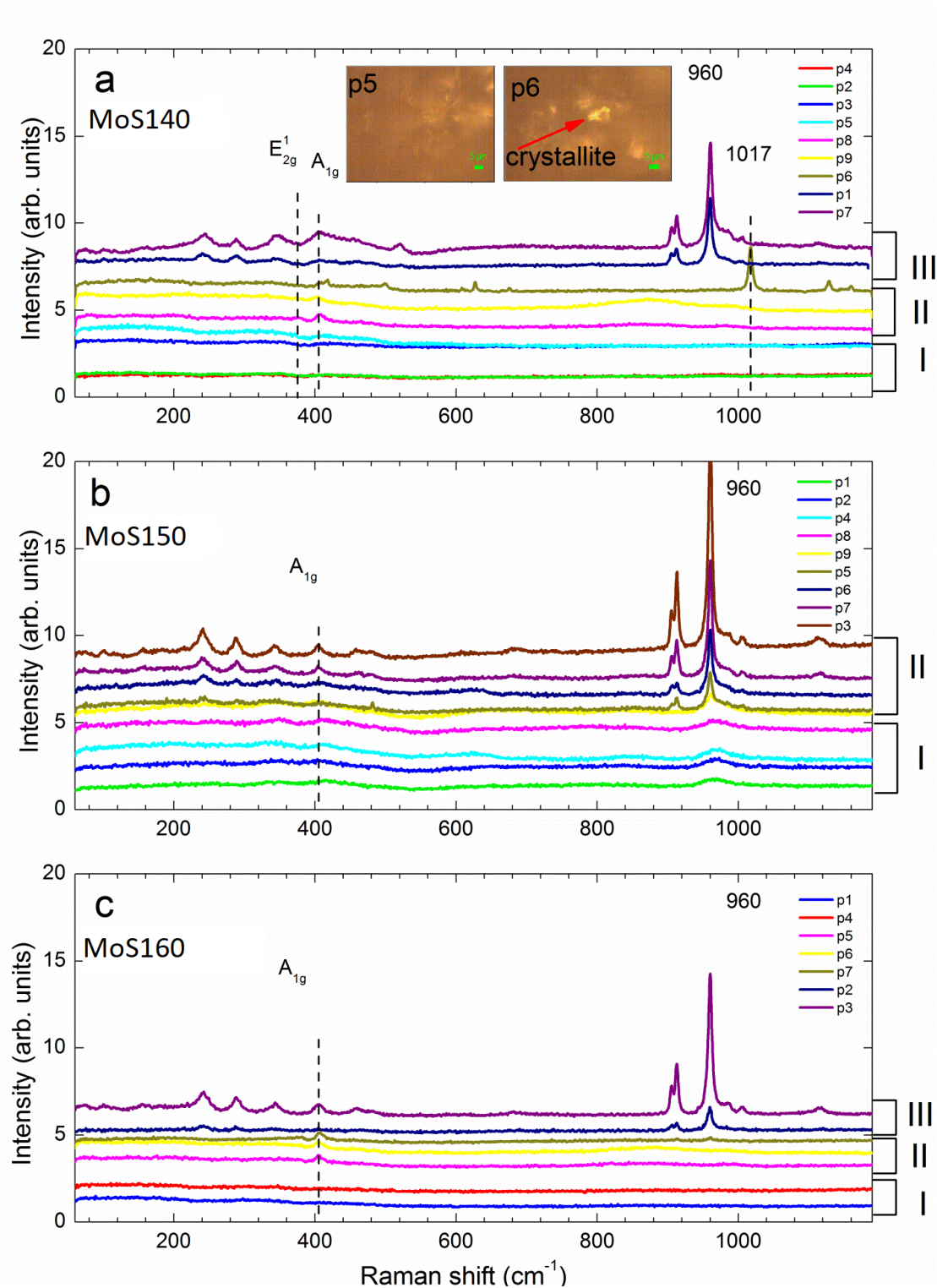

**Figure S5.** Raman spectra of the pressed MoS140O **(a)**, MoS150O **(b)** and MoSO160 **(c)** samples.

**Table T2.** The samples characterization from Raman data

| Sample | Amorphous phase | The MoS$_2$ phase | Crystalline oxide phase | Increase in excitation power |
|---|---|---|---|---|
| MoSO140 | (I) spectra of the amorphous phase with broad bands (p2, p3, p4, p5) | (II) spectra with MoS$_2$ phase bands characteristic of nanostructures (p8, p9) | (III) spectra with narrow bands of probably crystalline molybdenum oxide (p1, p6, p7). Spectra of type (III) are registered for areas in the form of crystallites of regular shape. | Oxidation (or crystallization) to the α-MoO$_3$ phase occurs for the amorphous areas, which is evidenced by the appearance of narrow intense bands. For the areas showing narrow Raman bands of oxides at low excitation power, their amorphization occurs under the power increase, as evidenced by the broadening of the bands |
| MoSO150 | (I) spectra of the amorphous phase with broad bands (p1, p2, p4, p8) | absent | (II) spectra with narrow bands of molybdenum oxide (not α-MoO$_3$) (p3, p5, p6, p7, p9). Spectra of type (II) are recorded for areas in the form of crystallites of regular shape. | For amorphous areas, there is a slight restructuring of the spectrum, probably due to oxidation, the bands become more pronounced, but typical for the amorphous phase. For areas that had narrow bands at low power, their amorphization occurs when the power is increased, which is evidenced by the broadening of the bands. |
| MoSO160 | (I) spectra of the amorphous phase with | (II) spectra with MoS$_2$ phase bands characteristic | (III) spectra with narrow bands probably of crystalline oxide (not α-MoO$_3$) of | For amorphous areas, their oxidation (or crystallization) to the α-MoO$_3$ phase occurs, as evidenced by the appearance |

| | | | |
|---|---|---|---|
| broad bands (p1, p4) | of nanostructures (p5, p6, p7) | molybdenum (p2, p3). Spectra of type (III) are registered for areas in the form of crystallites of regular shape. | of corresponding narrow intense bands. For areas that had narrow bands at low power, the shape of the spectrum does not change drastically when the power is increased. |

## References for Supporting Information